\documentstyle[12pt]{article}
\begin{document}
\hfill hep-th/0603126
\begin{center}
{\Large \bf Super Picard-Fuchs Equation and Monodromies for
Supermanifolds} \vskip 0.1in {Payal Kaura\footnote{email:
pa123dph@iitr.ernet.in}, Aalok Misra\footnote{e-mail:
aalokfph@iitr.ernet.in}
and Pramod Shukla\footnote{email: pmathpph@iitr.ernet.in}\\
Indian Institute of Technology Roorkee,\\
Roorkee - 247 667, Uttaranchal, India}
\vskip 0.5 true in
\end{center}

\begin{abstract}
Following \cite{AV} and \cite{HV}, we discuss the Picard-Fuchs equation for the super Landau-Ginsburg mirror to the super-Calabi-Yau in ${\bf WCP}^{(3|2)}[1,1,1,3|1,5]$, (using techniques of \cite{GL,AM}) Meijer basis
of solutions and monodromies (at 0,1 and $\infty$) in the large and small complex structure limits, as well as
 obtain the mirror hypersurface, which in the large K\"{a}hler limit, turns
out to be either a bidegree-(6,6) hypersurface in ${\bf WCP}^{(3|1)}[1,1,1,2]
\times {\bf WCP}^{(1|1)}[1,1|6]$ or a (${\bf Z}_2$-singular) bidegree-(6,12)
hypersurface in ${\bf WCP}^{(3|1)}[1,1,2,6|6]\times{\bf WCP}^{(1|1)}[1,1|6]$.
\end{abstract}

\section{Introduction}

The periods are the building blocks, e.g., for getting the
prepotential (and hence the K\"{a}hler potential and the Yukawa
coupling) in ${\cal N}=2$ type $II$ theories compactified on a
Calabi-Yau, and the superpotential for ${\cal N}=1$ type $II$
compactifications in the presence of (RR) fluxes. It is in this
regard that the Picard-Fuchs equation satisfied by the periods,
become quite important. Also, traversing non-trivial loops in the
complex structure moduli space of type $IIB$ on a Calabi-Yau mirror
to the one on the type $IIA$ side, corresponds to shifting of the
K\"{a}hler moduli in the K\"{a}hler moduli space on the type $IIA$
side. This results in mixing of flux numbers corresponding to $RR$
fluxes on the type $IIA$ side, implying thereby that dimensions of
cycles on the type $IIA$ side, loose their meaning. The mixing
matrix for the flux numbers is given by the monodromy matrix.  It
hence becomes important to evaluate the same.

In the context of generalizing mirror symmetry to include rigid manifolds, it was conjectured in
\cite{Sethi} that the mirrors for the same are given by supermanifolds. Further, in the past couple of
years, supermanifolds have been shown to be relevant to open/closed string dualities \cite{Witten}.

In this paper, we study some algebraic geometric aspects of a
supermanifold in a super weighted complex projective space free of
any potential orbifold singularities. In section {\bf 1},
based on techniques developed
in \cite{AV} in the context of supermanifolds and \cite{GL,AM}
regarding evaluation of Meijer basis for the periods as solutions to
the Picard-Fuchs equation, we obtain the Super Picard-Fuchs equation
for the mirror to a super Calabi-Yau and the periods for the same,
both in the large and small complex structure limits. We further
discuss the monodromies at 0,1 and $\infty$ (using again techniques
developed in \cite{GL,AM}). In section {\bf 2}, we obtain
the mirror hyerpsurface involving supermanifolds which are not
super Calabi-Yau's.

\section{Super Picard-Fuchs Equation, Meijer basis for periods and
Monodromies}

In this section, we discuss the super Picard-Fuchs equation for the
super Landau-Ginsburg mirror to the super-Calabi-Yau in
${\bf WCP}^{(3|2)}[1,1,1,3|1,5]$, (using techniques of \cite{GL,AM})
Meijer basis of solutions and monodromies in the large and small
complex structure limits.

\subsection{The Periods}

As shown in \cite{AM}, the gauged linear sigma model corresponding
to the supermanifold ${\bf WCP}^{(3|2)}[1,1,1,3|1,5]$ consists of
four chiral superfields $\Phi^{I=0,1,2,3}$ and two fermionic
superfields $\Theta^{I=0,1}$ satisfying the D-term constraint:
$|\Phi^0|^2 + |\Phi^1|^2 + |\Phi^2|^2 + 3 |\Phi^3|^2 + |\Theta^0|^2
+ 5 |\Theta^1|^2=r$. The mirror Super Landau-Ginsburg model is given
in terms of four twisted chiral superfields $Y^{I=0,1,2,3}$ (mirror
to the four $\Phi^I$'s), two more twisted chiral superfields
$X^{I=0,1}$ (mirror to the two $\Theta^I$'s), and two sets of
fermionic superfields $\eta^{I=0,1}$ and $\chi^{I=0,1}$ satisfying
the mirror constraint: $Y^0 + Y^1 + Y^2 + 3 Y^3 - X^0 - 5 X^1 = t$.
The periods can be expressed as:
\begin{equation}
\label{eq:period1} \Pi(t)=\int \prod_{I=0}^3 dY^I \prod_{J=0}^1 dX^J
d\eta^J d\chi^J e^{\sum_{I=0}^3e^{-Y^I} + \sum_{J=0}^1e^{-X^J}(1 +
\eta^J\chi^J)}\delta(\sum_{I=0}^3Q^IY^I - \sum_{J=0}^1q^JX^J - t),
\end{equation}
where $Q^{0,1,2,3}\equiv(1,1,1,3)$ and $q^{0,1}\equiv(1,5)$.
In the spirit of \cite{HV}, consider now:
\begin{equation}
\label{eq:period2} \tilde{\Pi}(t)=\int \prod_{I=0}^3 dY^I
\prod_{J=0}^1 dX^J d\eta^J d\chi^J e^{\sum_{I=0}^3\mu_I e^{-Y^I} +
\sum_{J=0}^1\nu_Je^{-X^J} +
\sum_{J=0}^1e^{-X^J}\eta^J\chi^J}\delta(\sum_{I=0}^3Q^IY^I -
\sum_{J=0}^1q^JX^J - t).
\end{equation}
The deformations $\mu^I$'s and $\nu^J$'s can be absorbed into shifting the K\"{a}hler parameter $t$ to
$t^\prime = t - ln\biggl(\frac{\mu_0\mu_1\mu_2\mu_3^3}{\nu_0\nu_1^5}\biggr)$. One can then see that one
gets the following Super Picard-Fuchs equation:
\begin{equation}
\label{eq:SP1F}
\frac{\partial^6\tilde{\Pi}(t^\prime)}{\partial\mu_0\partial\mu_1\partial\mu_2\partial\mu_3^3}=e^{-t^\prime}
\frac{\partial^6\tilde{\Pi}(t^\prime)}{\partial\nu_0\partial\nu_1^5}.
\end{equation}
By noticing:
\begin{equation}
\label{eq:identity}
\frac{\partial}{\partial\mu_I^{Q^I}}=\prod_{i=0}^{Q^I-1}(-Q^I\frac{d}{dt^\prime}-i),\
\frac{\partial}{\partial\nu_J^{q^J}}=\prod_{i=0}^{q^J-1}(q^J\frac{d}{dt^\prime}-i),
\end{equation}
one gets the following:
\begin{equation}
\label{eq:SPF2}
\biggl(-\frac{d}{dt^\prime}\biggr)^3\prod_{i=0}^2\biggl(-3\frac{d}{dt^\prime}-i\biggr)\tilde{\Pi}(t^\prime)
= e^{-t^\prime}\frac{d}{dt^\prime}\prod_{i=0}^4\biggl(5\frac{d}{dt^\prime}-i\biggr)\tilde{\Pi}(t^\prime).
\end{equation}
Further, setting $e^{-t^\prime}\equiv z$, and by replacing $\frac{5^5}{3^3}z$ by $z$ (noticing that
$\Delta_z\equiv z\frac{d}{dz}$ is scale invariant), one gets the final form of the Picard-Fuchs equation:
\begin{equation}
\label{eq:SPF3}
\Delta_z^4(\Delta_z - \frac{1}{3})(\Delta_z - \frac{2}{3})\Pi(z) = z \Delta_z^2\prod_{i=1}^4(\Delta_z + \frac{i}{5})\tilde\Pi(z).
\end{equation}
Comparing \ref{eq:SPF3} with the following equation for the generalized hypergeometric functions:
\begin{equation}
\label{eq:pFq} \biggl[\Delta_z\prod_{i=1}^q(\Delta_z + \beta_i - 1)
- z\prod_{j=1}^p(\Delta_z + \alpha_j)\biggr]\tilde\Pi(z)=0,
\end{equation}
the solution to which is given by
$\ _6F_5\left(\begin{array}{cccccc} 0 & 0 & \frac{1}{5} & \frac{2}{5} & \frac{3}{5} & \frac{4}{5} \\
1 & 1 & 1 & \frac{2}{3} & \frac{1}{3}& \\
\end{array}\right)(z)$.

From the above solution, following \cite{GL}, the Meijer basis of solutions is obtained using properties of
$\ _pF_q$ and the Meijer function $I$:
\begin{eqnarray}
\label{eq:IpFqprops}
& & \ _pF_q\left(\begin{array}{cccc}
\alpha_1 & \alpha_2 & \alpha_3 & ....\ \alpha_p\\
\beta_1 & \beta_2 & \beta_3 & ....\ \beta_q\\
\end{array}\right)(z)={\prod_{i=1}^p\Gamma(\beta_i)\over\prod_{j=1}^q\Gamma(\alpha_j)}
I\left(\begin{array}{c|c}
0 & \alpha_1...\alpha_p\\
\hline
. & \beta_1...\beta_q\\
\end{array}\right)(-z)\ {\rm where}\nonumber\\
& & I\left(\begin{array}{c|c}
a_1...a_A & b_1...b_B\\
\hline
c_1...c_C & d_1...d_D\\
\end{array}\right)(z), I\left(\begin{array}{c|c}
a_1...(1-d_l)...a_A & b_1...b_B\\
\hline
c_1...c_C & d_1...\hat{d}_l...d_D\\
\end{array}\right)(-z)\nonumber\\
& & I\left(\begin{array}{c|c}
a_1...a_A & b_1..\hat{b}_j...b_B\\
\hline
c_1..(1-b_j)..c_C & d_1...d_D\\
\end{array}\right)(-z),
\end{eqnarray}
where a hat implies that the corresponding entry is missing, satisfy the same equation.
Now, one would mimick the symplectic structure for bosonic manifolds, for supermanifolds as well, and
construct the following column period vector:
\begin{equation}
\label{eq:sympper}
\Pi(z)=\left(\begin{array}{c}F_0\\F_1\\F_2\\Z^0\\Z^1\\Z^2\\
\end{array}\right).
\end{equation}
Now, to get an infinite series expansion in $z$ for $|z|<1$ as well as $|z|>1$,
one uses the following
\begin{equation}
\label{eq:MBdieuf}
I\left(\begin{array}{c|c}
a_1...a_A & b_1...b_B\\ \hline
c_1...c_C & d_1...d_D\\
\end{array}\right)(z)={1\over2\pi i}\int_\gamma ds {\prod_{i=1}^A\Gamma(a_i-s)\prod_{j=1}^B
\Gamma(b_j+s)\over\prod_{k=1}^C\Gamma(c_k-s)\prod_{l=1}^D\Gamma(d_l+s)}z^s,
\end{equation}
where the contour $\gamma$ lies to the right
of:$s+b_j=-m\in{\bf Z}^-\cup\{0\}$ and
to the left of: $a_i-s=-m\in{\bf Z}^-\cup\{0\}$.

This, $|z|<<1$ and $|z|>>1$ can be dealt with equal ease by suitable deformations of the contour (See \cite{GL,AM}).
Additionally, instead of performing parametric differentiation of
infinite series to get the  $ln$-terms, one get the same (for the large complex structure
limit: $|z|<1$) by evaluation of the residue at $s=0$ in the Mellin-Barnes contour integral
in (\ref{eq:MBdieuf}) as is done explicitly to evaluate the six integrals.

The guiding principle is that of the six solutions to ${\Pi}$, one should
generate solutions in which one gets $(ln z)^P$, $P=0,...,3$ and one can then identify terms independent of $ln z$ with
$Z^0$, three $(ln z)$ terms with $Z^{1,2,3}$, three $(ln z)^{P\leq2}$
terms with $F_{1,2,3}\equiv
{\partial F\over\partial Z^{1,2,3}}$,
and finally $(ln z)^{P\leq3}$ term with $F_0\equiv{\partial F\over\partial Z^0}$.

One (non-unique) choice of solutions for ${\Pi}(z)$ is given below:
\begin{eqnarray}
\label{eq:sols}
\left(\begin{array}{c}
I\left(\begin{array}{c|c}
0\ 0 & 0\ 0\ {1\over5}\ {2\over5}\ {3\over5}\  {4\over5}\\
& \\ \hline
. & 1\ 1\ {2\over3}\ {1\over3}\\
\end{array}\right)(z)\\
\hfill\\
I\left(\begin{array}{c|c}
0 & 0\ 0\ {1\over5}\ {2\over5}\ {3\over5}\  {4\over5}\\
& \\ \hline
. & 1\ 1\ 1\ {2\over3}\ {1\over3}\\
\end{array}\right)(-z)\\
\hfill\\
I\left(\begin{array}{c|c}
 0 & 0\ 0\  {2\over5}\ {3\over5}\  {4\over5}\\
& \\ \hline
{4\over5} & 1\ 1\ 1\ {2\over3}\ {1\over3}\\
\end{array}\right)(z)\\
\hfill\\
I\left(\begin{array}{c|c}
0 &  {1\over5}\ {2\over5}\ {3\over5}\  {4\over5}\\
& \\ \hline
1\ 1 & 1\ 1\ 1\ {2\over3}\ {1\over3}\\
\end{array}\right)(-z)\\
\hfill\\
I\left(\begin{array}{c|c}
0 &  0\ {2\over5}\ {3\over5}\  {4\over5}\\
& \\ \hline
1\ {4\over5} & 1\ 1\ 1\ {2\over3}\ {1\over3}\\
\end{array}\right)(-z)\\
\hfill\\
I\left(\begin{array}{c|c}
0 & 0\  {3\over5}\  {4\over5}\\
& \\ \hline
1\ {4\over5}\ {3\over5} & 1\ 1\ 1\ {2\over3}\ {1\over3}\\
\end{array}\right)(-z)\\
\end{array}\right)
=\left[\begin{array}{c}
F_0\\
F_1\\
F_2\\
Z^0\\
Z^1\\
Z^2\\
\end{array}\right]
\end{eqnarray}

Using techniques of \cite{GL,AM}, and defining (the polygamma
function)  $\psi(z)\equiv\frac{\Gamma^\prime(z)}{\Gamma(z)}$, one
gets the following results:
\begin{eqnarray}
\label{eq:F_0}
& & F_0=\biggl[
\frac{{\Gamma(-( \frac{1}{5} ) )}^2\,
    {\Gamma(\frac{1}{5})}^3\,
    \Gamma(\frac{2}{5})\,
    \Gamma(\frac{3}{5})}{z^{\frac{1}{5}}\,
    \Gamma(\frac{2}{15})\,
    \Gamma(\frac{7}{15})\,
    {\Gamma(\frac{4}{5})}^2}
+ \frac{{\Gamma(-( \frac{2}{5} ) )}^2\,
    \Gamma(-( \frac{1}{5} ) )\,
    \Gamma(\frac{1}{5})\,
    {\Gamma(\frac{2}{5})}^3}{z^{\frac{2}{5}}\,
    \Gamma(-( \frac{1}{15} ) )\,
    \Gamma(\frac{4}{15})\,
    {\Gamma(\frac{3}{5})}^2}
+ \frac{{\Gamma(-( \frac{3}{5} ) )}^2\,
    \Gamma(-( \frac{2}{5} ) )\,
    \Gamma(-( \frac{1}{5} ) )\,
    \Gamma(\frac{1}{5})\,
    {\Gamma(\frac{3}{5})}^2}{z^{\frac{3}{5}}\,
    \Gamma(-( \frac{4}{15} ) )\,
    \Gamma(\frac{1}{15})\,
    {\Gamma(\frac{2}{5})}^2}\nonumber\\
& & +
\frac{{\Gamma(-( \frac{4}{5} ) )}^2\,
    \Gamma(-( \frac{3}{5} ) )\,
    \Gamma(-( \frac{2}{5} ) )\,
    \Gamma(-( \frac{1}{5} ) )\,
    {\Gamma(\frac{4}{5})}^2}{z^{\frac{4}{5}}\,
    \Gamma(-( \frac{7}{15} ) )\,
    \Gamma(-( \frac{2}{15} ) )\,
    {\Gamma(\frac{1}{5})}^2}
-( \frac{{\Gamma(-( \frac{6}{5} )
            )}^2\,\Gamma(-( \frac{4}{5} )
          )\,\Gamma(-( \frac{3}{5} ) )\,
      \Gamma(-( \frac{2}{5} ) )\,
      {\Gamma(\frac{6}{5})}^2}{z^{\frac{6}{5}}\,
      \Gamma(-( \frac{13}{15} ) )\,
      \Gamma(-( \frac{8}{15} ) )\,
      {\Gamma(-( \frac{1}{5} ) )}^2}
    )\nonumber\\
& &
-( \frac{{\Gamma(-( \frac{7}{5} )
            )}^2\,\Gamma(-( \frac{6}{5} )
          )\,\Gamma(-( \frac{4}{5} ) )\,
      \Gamma(-( \frac{3}{5} ) )\,
      {\Gamma(\frac{7}{5})}^2}{z^{\frac{7}{5}}\,
      \Gamma(-( \frac{16}{15} ) )\,
      \Gamma(-( \frac{11}{15} ) )\,
      {\Gamma(-( \frac{2}{5} ) )}^2}
    )
-( \frac{{\Gamma(-( \frac{8}{5} )
            )}^2\,\Gamma(-( \frac{7}{5} )
          )\,\Gamma(-( \frac{6}{5} ) )\,
      \Gamma(-( \frac{4}{5} ) )\,
      {\Gamma(\frac{8}{5})}^2}{z^{\frac{8}{5}}\,
      \Gamma(-( \frac{19}{15} ) )\,
      \Gamma(-( \frac{14}{15} ) )\,
      {\Gamma(-( \frac{3}{5} ) )}^2}
    )\nonumber\\
& & -( \frac{{\Gamma(-( \frac{9}{5} )
            )}^2\,\Gamma(-( \frac{8}{5} )
          )\,\Gamma(-( \frac{7}{5} ) )\,
      \Gamma(-( \frac{6}{5} ) )\,
      {\Gamma(\frac{9}{5})}^2}{z^{\frac{9}{5}}\,
      \Gamma(-( \frac{22}{15} ) )\,
      \Gamma(-( \frac{17}{15} ) )\,
      {\Gamma(-( \frac{4}{5} ) )}^2}
    )+...\biggr]\theta(|z|-1)\nonumber\\
& & \!\!\!\!\!\!\!\!+\theta(1-|z|)\biggl[ \frac{{\cal X}} {6\,
    \Gamma(\frac{1}{3})\,
    \Gamma(\frac{2}{3})} +\frac{z\,\Gamma(\frac{6}{5})\,
    \Gamma(\frac{7}{5})\,
    \Gamma(\frac{8}{5})\,
    \Gamma(\frac{9}{5})\,
    ( -4 + 2\,\gamma + \log (z) +
      \psi\frac{6}{5}) -
      \psi(\frac{4}{3}) +
      \psi(\frac{7}{5}) +
      \psi(\frac{8}{5}) -
      \psi(\frac{5}{3}) +
      \psi(\frac{9}{5}) ) }{
    \Gamma(\frac{4}{3})\,
    \Gamma(\frac{5}{3})}\nonumber\\
& & \frac{z^2\,\Gamma(\frac{11}{5})\,
    \Gamma(\frac{12}{5})\,
    \Gamma(\frac{13}{5})\,
    \Gamma(\frac{14}{5})\,
    ( -4 + 2\,\gamma + \log (z) +
      \psi(\frac{11}{5}) -
      \psi(\frac{7}{3}) +
      \psi\frac{12}{5}) +
      \psi(\frac{13}{5}) -
      \psi(\frac{8}{3}) +
      \psi(\frac{14}{5}) ) }{16\,
    \Gamma(\frac{7}{3})\,
    \Gamma(\frac{8}{3})}+...\biggr],\nonumber\\
& &
\end{eqnarray}
where
$${\cal X}\equiv \Gamma(\frac{1}{5})\,
    \Gamma(\frac{2}{5})\,
    \Gamma(\frac{3}{5})\,
    \Gamma(\frac{4}{5})\,
    ( 8\,{\gamma}^3 + ({\log (z)})^3 +
      {\pi }^2\,\psi(\frac{1}{5}) +
      {\psi(\frac{1}{5})}^3-
      {\pi }^2\,\psi(\frac{1}{3}) -
      3\,{\psi(\frac{1}{5})}^2\,
       \psi(\frac{1}{3}) +
      3\,\psi(\frac{1}{5})\,
       {\psi(\frac{1}{3})}^2 -
      {\psi(\frac{1}{3})}^3\hfill$$
$$ +
      {\pi }^2\,\psi(\frac{2}{5}) +
      3\,{\psi(\frac{1}{5})}^2\,
       \psi(\frac{2}{5}) -
      6\,\psi(\frac{1}{5})\,
       \psi(\frac{1}{3})\,
       \psi(\frac{2}{5}) +
      3\,{\psi(\frac{1}{3})}^2\,
       \psi(\frac{2}{5})+
      3\,\psi(\frac{1}{5})\,
       {\psi(\frac{2}{5})}^2 -
      3\,\psi(\frac{1}{3})\,
       {\psi(\frac{2}{5})}^2 +
      {\psi(\frac{2}{5})}^3 +
      {\pi }^2\,\psi(\frac{3}{5})\hfill$$
      $$\!\!\!\!\!\!\!\! +
      3\,{\psi(\frac{1}{5})}^2\,
       \psi(\frac{3}{5})-
      6\,\psi(\frac{1}{5})\,
       \psi(\frac{1}{3})\,
       \psi(\frac{3}{5}) +
      3\,{\psi(\frac{1}{3})}^2\,
       \psi(\frac{3}{5}) +
      6\,\psi(\frac{1}{5})\,
       \psi(\frac{2}{5})\,
       \psi(\frac{3}{5}) -
      6\,\psi(\frac{1}{3})\,
       \psi(\frac{2}{5})\,
       \psi(\frac{3}{5}) +
      3\,{\psi(\frac{2}{5})}^2\,
       \psi(\frac{3}{5}) +
      3\,\psi(\frac{1}{5})\,
       {\psi(\frac{3}{5})}^2\hfill$$
$$\!\!\! -
      3\,\psi(\frac{1}{3})\,
       {\psi(\frac{3}{5})}^2 +
      3\,\psi(\frac{2}{5})\,
       {\psi(\frac{3}{5})}^2 +
      {\psi(\frac{3}{5})}^3 -
      {\pi }^2\,\psi(\frac{2}{3}) -
      3\,{\psi(\frac{1}{5})}^2\,
       \psi(\frac{2}{3})+
      6\,\psi(\frac{1}{5})\,
       \psi(\frac{1}{3})\,
       \psi(\frac{2}{3}) -
      3\,{\psi(\frac{1}{3})}^2\,
       \psi(\frac{2}{3}) -
      6\,\psi(\frac{1}{5})\,
       \psi(\frac{2}{5})\,
       \psi(\frac{2}{3})\hfill$$
$$ +
      6\,\psi(\frac{1}{3})\,
       \psi(\frac{2}{5})\,
       \psi(\frac{2}{3}) -
      3\,{\psi(\frac{2}{5})}^2\,
       \psi(\frac{2}{3}) -
      6\,\psi(\frac{1}{5})\,
       \psi(\frac{3}{5})\,
       \psi(\frac{2}{3})+
      6\,\psi(\frac{1}{3})\,
       \psi(\frac{3}{5})\,
       \psi(\frac{2}{3}) -
      6\,\psi(\frac{2}{5})\,
       \psi(\frac{3}{5})\,
       \psi(\frac{2}{3}) -
      3\,{\psi(\frac{3}{5})}^2\,
       \psi(\frac{2}{3})\hfill$$
       $$ +
      3\,\psi(\frac{1}{5})\,
       {\psi(\frac{2}{3})}^2 -
      3\,\psi(\frac{1}{3})\,
       {\psi(\frac{2}{3})}^2 +
      3\,\psi(\frac{2}{5})\,
       {\psi(\frac{2}{3})}^2 +
      3\,\psi(\frac{3}{5})\,
       {\psi(\frac{2}{3})}^2 -
      {\psi(\frac{2}{3})}^3 +
      {\pi }^2\,\psi(\frac{4}{5})
+
      3\,{\psi(\frac{1}{5})}^2\,
       \psi(\frac{4}{5}) -
      6\,\psi(\frac{1}{5})\,
       \psi(\frac{1}{3})\,
       \psi(\frac{4}{5})\hfill$$
$$ +
      3\,{\psi(\frac{1}{3})}^2\,
       \psi(\frac{4}{5}) +
      6\,\psi(\frac{1}{5})\,
       \psi(\frac{2}{5})\,
       \psi(\frac{4}{5}) -
      6\,\psi(\frac{1}{3})\,
       \psi(\frac{2}{5})\,
       \psi(\frac{4}{5}) +
      3\,{\psi(\frac{2}{5})}^2\,
       \psi(\frac{4}{5}) 6\,\psi(\frac{1}{5})\,
       \psi(\frac{3}{5})\,
       \psi(\frac{4}{5})-
      6\,\psi(\frac{1}{3})\,
       \psi(\frac{3}{5})\,
       \psi(\frac{4}{5}) \hfill$$
$$+
      6\,\psi(\frac{2}{5})\,
       \psi(\frac{3}{5})\,
       \psi(\frac{4}{5}) +
      3\,{\psi(\frac{3}{5})}^2\,
       \psi(\frac{4}{5}) -
      6\,\psi(\frac{1}{5})\,
       \psi(\frac{2}{3})\,
       \psi(\frac{4}{5}) +
      6\,\psi(\frac{1}{3})\,
       \psi(\frac{2}{3})\,
       \psi(\frac{4}{5})  -
      6\,\psi(\frac{2}{5})\,
       \psi(\frac{2}{3})\,
       \psi(\frac{4}{5}) -
      6\,\psi(\frac{3}{5})\,
       \psi(\frac{2}{3})\,
       \psi(\frac{4}{5})\hfill$$
$$ +
      3\,{\psi(\frac{2}{3})}^2\,
       \psi(\frac{4}{5}) +
      3\,\psi(\frac{1}{5})\,
       {\psi(\frac{4}{5})}^2  -
      3\,\psi(\frac{1}{3})\,
       {\psi(\frac{4}{5})}^2 +
      3\,\psi(\frac{2}{5})\,
       {\psi(\frac{4}{5})}^2 +
      3\,\psi(\frac{3}{5})\,
       {\psi(\frac{4}{5})}^2 -
      3\,\psi(\frac{2}{3})\,
       {\psi(\frac{4}{5})}^2 +
      {\psi(\frac{4}{5})}^3 +
      12\,{\gamma}^2\,
       ( \psi(\frac{1}{5})\hfill$$
$$ -
         \psi(\frac{1}{3}) +
         \psi(\frac{2}{5}) +
         \psi(\frac{3}{5}) -
         \psi(\frac{2}{3}) +
         \psi(\frac{4}{5}) )  +
      3\,({\log (z)})^2\,( 2\,\gamma +
         \psi(\frac{1}{5})-
         \psi(\frac{1}{3}) +
         \psi(\frac{2}{5}) +
         \psi(\frac{3}{5}) -
         \psi(\frac{2}{3}) +
         \psi(\frac{4}{5}) )  +
      3\,\psi(\frac{1}{5})\,
       \psi^\prime(\frac{1}{5}) \hfill$$
       $$\!\!\!-
      3\,\psi(\frac{1}{3})\,
       \psi^\prime(\frac{1}{5})+
      3\,\psi(\frac{2}{5})\,
       \psi^\prime(\frac{1}{5}) +
      3\,\psi(\frac{3}{5})\,
       \psi^\prime(\frac{1}{5}) -
      3\,\psi(\frac{2}{3})\,
       \psi^\prime(\frac{1}{5}) +
      3\,\psi(\frac{4}{5})\,
       \psi^\prime(\frac{1}{5}) -
      3\,\psi(\frac{1}{5})\,
       \psi^\prime(\frac{1}{3})+
      3\,\psi(\frac{1}{3})\,
       \psi^\prime(\frac{1}{3}) -
      3\,\psi(\frac{2}{5})\,
       \psi^\prime(\frac{1}{3})\hfill$$
       $$ \!\!\!\!-
      3\,\psi(\frac{3}{5})\,
       \psi^\prime(\frac{1}{3}) +
      3\,\psi(\frac{2}{3})\,
       \psi^\prime(\frac{1}{3})-
      3\,\psi(\frac{4}{5})\,
       \psi^\prime(\frac{1}{3}) +
      3\,\psi(\frac{1}{5})\,
       \psi^\prime(\frac{2}{5}) -
      3\,\psi(\frac{1}{3})\,
       \psi^\prime(\frac{2}{5}) +
      3\,\psi(\frac{2}{5})\,
       \psi^\prime(\frac{2}{5})+
      3\,\psi(\frac{3}{5})\,
       \psi^\prime(\frac{2}{5}) -
      3\,\psi(\frac{2}{3})\,
       \psi^\prime(\frac{2}{5})\hfill$$
       $$ \!\!\!+
      3\,\psi(\frac{4}{5})\,
       \psi^\prime(\frac{2}{5}) +
      3\,\psi(\frac{1}{5})\,
       \psi^\prime(\frac{3}{5})-
      3\,\psi(\frac{1}{3})\,
       \psi^\prime(\frac{3}{5}) +
      3\,\psi(\frac{2}{5})\,
       \psi^\prime(\frac{3}{5}) +
      3\,\psi(\frac{3}{5})\,
       \psi^\prime(\frac{3}{5}) -
      3\,\psi(\frac{2}{3})\,
       \psi^\prime(\frac{3}{5}) +
      3\,\psi(\frac{4}{5})\,
       \psi^\prime(\frac{3}{5})-
      3\,\psi(\frac{1}{5})\,
       \psi^\prime(\frac{2}{3})\hfill$$
       $$ +
      3\,\psi(\frac{1}{3})\,
       \psi^\prime(\frac{2}{3}) -
      3\,\psi(\frac{2}{5})\,
       \psi^\prime(\frac{2}{3}) -
      3\,\psi(\frac{3}{5})\,
       \psi^\prime(\frac{2}{3}) +
      3\,\psi(\frac{2}{3})\,
       \psi^\prime(\frac{2}{3})-
      3\,\psi(\frac{4}{5})\,
       \psi^\prime(\frac{2}{3}) +
      3\,\psi(\frac{1}{5})\,
       \psi^\prime(\frac{4}{5}) -
      3\,\psi(\frac{1}{3})\,
       \psi^\prime(\frac{4}{5}) \hfill$$
$$+
      3\,\psi(\frac{2}{5})\,
       \psi^\prime(\frac{4}{5}) +
      3\,\psi(\frac{3}{5})\,
       \psi^\prime(\frac{4}{5})-
      3\,\psi(\frac{2}{3})\,
       \psi^\prime(\frac{4}{5}) +
      3\,\psi(\frac{4}{5})\,
       \psi^\prime(\frac{4}{5}) +
      2\,\gamma\,
       ( {\pi }^2 + 3\,
          ( {\psi(\frac{1}{5})}^2 +
            {\psi(\frac{1}{3})}^2 +
            {\psi(\frac{2}{5})}^2\hfill$$
$$ +
            2\,\psi(\frac{2}{5})\,
             \psi(\frac{3}{5}) +
            {\psi(\frac{3}{5})}^2 -
            2\,\psi(\frac{2}{5})\,
             \psi(\frac{2}{3}) -
            2\,\psi(\frac{3}{5})\,
             \psi(\frac{2}{3}) +
            {\psi(\frac{2}{3})}^2 -
            2\,\psi(\frac{1}{5})\,
             ( \psi(\frac{1}{3})-
               \psi(\frac{2}{5}) -
               \psi(\frac{3}{5}) +
               \psi(\frac{2}{3})\hfill$$
               $$ -
               \psi(\frac{4}{5}) )  +
            2\,\psi(\frac{2}{5})\,
             \psi(\frac{4}{5}) +
            2\,\psi(\frac{3}{5})\,
             \psi(\frac{4}{5}) -
            2\,\psi(\frac{2}{3})\,
             \psi(\frac{4}{5}) +
            {\psi(\frac{4}{5})}^2-
            2\,\psi(\frac{1}{3})\,
             ( \psi(\frac{2}{5}) +
               \psi(\frac{3}{5}) -
               \psi(\frac{2}{3}) +
               \psi(\frac{4}{5}) )  +
            \psi^\prime(\frac{1}{5}) -
            \psi^\prime(\frac{1}{3})\hfill$$
$$ \!\!\!\!+
            \psi^\prime(\frac{2}{5}) +
            \psi^\prime(\frac{3}{5}) -
            \psi^\prime(\frac{2}{3}) +
            \psi^\prime(\frac{4}{5}) )
         )  + \log (z)\,
       ( 12\,{\gamma}^2 + {\pi }^2 +
         12\,\gamma\,
          ( \psi(\frac{1}{5}) -
            \psi(\frac{1}{3})+
            \psi(\frac{2}{5}) +
            \psi(\frac{3}{5}) -
            \psi(\frac{2}{3}) +
            \psi(\frac{4}{5}) )  +
         3\,( {\psi(\frac{1}{5})}^2 +
            {\psi(\frac{1}{3})}^2\hfill$$
$$ +
            {\psi(\frac{2}{5})}^2 +
            2\,\psi(\frac{2}{5})\,
             \psi(\frac{3}{5}) +
            {\psi(\frac{3}{5})}^2 -
            2\,\psi(\frac{2}{5})\,
             \psi(\frac{2}{3}) -
            2\,\psi(\frac{3}{5})\,
             \psi(\frac{2}{3})+
            {\psi(\frac{2}{3})}^2 -
            2\,\psi(\frac{1}{5})\,
             ( \psi(\frac{1}{3}) -
               \psi(\frac{2}{5}) -
               \psi(\frac{3}{5}) +
               \psi(\frac{2}{3}) -
               \psi(\frac{4}{5}) )\hfill$$
$$  +
            2\,\psi(\frac{2}{5})\,
             \psi(\frac{4}{5}) +
            2\,\psi(\frac{3}{5})\,
             \psi(\frac{4}{5}) -
            2\,\psi(\frac{2}{3})\,
             \psi(\frac{4}{5}) +
            {\psi(\frac{4}{5})}^2-
            2\,\psi(\frac{1}{3})\,
             ( \psi(\frac{2}{5}) +
               \psi(\frac{3}{5}) -
               \psi(\frac{2}{3}) +
               \psi(\frac{4}{5}) )  +
            \psi^\prime(\frac{1}{5})\hfill$$
$$ -
            \psi^\prime(\frac{1}{3}) +
            \psi^\prime(\frac{2}{5}) +
            \psi^\prime(\frac{3}{5}) -
            \psi^\prime(\frac{2}{3}) +
            \psi^\prime(\frac{4}{5}) )
         )  + \psi^{\prime\prime}(\frac{1}{5})-
      \psi^{\prime\prime}(\frac{2}{3}) +
      \psi^{\prime\prime}(\frac{2}{5}) +
      \psi^{\prime\prime}(\frac{3}{5}) -
      \psi^{\prime\prime}(\frac{2}{3}) +
      \psi^{\prime\prime}(\frac{4}{5}) -
      2\,\psi^{\prime\prime}(1) );$$
It would be beneficial to understand how the above result (which is the
most involved among all the periods) has been obtained. From the Mellin-Barnes
integral representation of the function relevant to the $F_0$'s evaluation,
one sees that one has to evaluate the following contour integral:
\begin{equation}
\label{eq:F0int1}
F_0=\frac{1}{2\pi i}\int_\gamma \frac{[\Gamma(-s)]^2[\Gamma(s)]^2
\prod_{j=1}^5\Gamma(\frac{j}{5} + s)}{[\Gamma(1+s)]^2\prod_{i=1}^2\Gamma
(\frac{i}{3}+s)}z^s ds.
\end{equation}
The above contour integral can be evaluated using the method of residues as
follows. One notices that the integrand has poles of order 4 at $s=0$
(relevant to $|z|<<1$),
of order 2 at $s=m\in{\bf Z}^+$ and of order 1 at
$s+\frac{j}{5}=-m\in{\bf Z}^-$ (relevant to $|z|>>1$).

To evaluate the residue at $s=0$, define:
\begin{equation}
\label{eq:F0int2}
\Omega_1(s)\equiv s^4 \frac{[\Gamma(-s)]^2[\Gamma(s)]^2
\prod_{j=1}^5\Gamma(\frac{j}{5} + s)}{[\Gamma(1+s)]^2\prod_{i=1}^2\Gamma
(\frac{i}{3}+s)}z^s.
\end{equation}
To evaluate the residue, one needs to evaluate $\frac{1}{6}\frac{d^3}{ds^3}
\Omega_1(s)|_{s=0}$. Taking the derivative of the logarithm of $\Omega_1(s)$,
one gets:
\begin{equation}
\label{eq:F0int3}
\Omega^\prime(s)=\Omega_1(s)\Omega_2(s),
\end{equation}
where $\Omega_2(s)\equiv -\Psi(1-s) + \sum_{j=1}^5\Psi(\frac{j}{5}+s)
-\sum_{i=1}^2\Psi(\frac{i}{3} + s) + ln z$. Similarly,using (\ref{eq:F0int3})
\begin{equation}
\label{eq:F0int4}
\Omega^{\prime\prime}_1(s)=\Omega_1(s)[\Omega_2(s)]^2
+ \Omega_1(s)\Omega^\prime_2(s),
\end{equation}
where $\Omega^\prime_2(s)\equiv \Psi^\prime(1-s) + \sum_{j=1}^5\Psi^\prime
(\frac{j}{5}+s)-\sum_{i=1}^2\Psi^\prime(\frac{i}{3}+s)$. Finally, using
(\ref{eq:F0int3}), one gets:
\begin{equation}
\label{eq:F0int5}
\Omega^{\prime\prime\prime}_1=\Omega_1(s)[\Omega_2(s)]^3+3\Omega_1(s)\Omega_2(s)
\Omega^\prime_2(s) + \Omega_1(s)\Omega_2^{\prime\prime}(s),
\end{equation}
where $\Omega^{\prime\prime}_2(s)=\Psi^{\prime\prime}(1-s) + \sum_{j=1}^5
\Psi^{\prime\prime}(\frac{j}{5}+s) - \sum_{i=1}^2\Psi^{\prime\prime}
(\frac{j}{3}+s)$. Putting everything together and expanding out the terms,
one gets (\ref{eq:F_0}).

One can similarly evaluate the other components of the period vector, which
are given below:
\begin{eqnarray}
\label{eq:F_1}
& & F_1=
\biggl[\frac{{\Gamma(-( \frac{1}{5} ) )}^2\,
    {\Gamma(\frac{1}{5})}^2\,
    \Gamma(\frac{2}{5})\,
    \Gamma(\frac{3}{5})}{{( -z ) }^
     {\frac{1}{5}}\,\Gamma(\frac{2}{15})\,
    \Gamma(\frac{7}{15})\,
    {\Gamma(\frac{4}{5})}^3} -
( \frac{{\Gamma(-( \frac{6}{5} )
            )}^2\,\Gamma(-( \frac{4}{5} )
          )\,\Gamma(-( \frac{3}{5} ) )\,
      \Gamma(-( \frac{2}{5} ) )\,
      \Gamma(\frac{6}{5})}{{( -z ) }^
       {\frac{6}{5}}\,\Gamma(-( \frac{13}
          {15} ) )\,
      \Gamma(-( \frac{8}{15} ) )\,
      {\Gamma(-( \frac{1}{5} ) )}^3}
    ) + \frac{{\Gamma(-( \frac{2}{5} ) )}^2\,
    \Gamma(-( \frac{1}{5} ) )\,
    \Gamma(\frac{1}{5})\,
    {\Gamma(\frac{2}{5})}^2}{{( -z ) }^
     {\frac{2}{5}}\,\Gamma(-( \frac{1}{15}
        ) )\,\Gamma(\frac{4}{15})\,
    {\Gamma(\frac{3}{5})}^3}\nonumber\\
& & +
-( \frac{{\Gamma(-( \frac{7}{5} )
            )}^2\,\Gamma(-( \frac{6}{5} )
          )\,\Gamma(-( \frac{4}{5} ) )\,
      \Gamma(-( \frac{3}{5} ) )\,
      \Gamma(\frac{7}{5})}{{( -z ) }^
       {\frac{7}{5}}\,\Gamma(-( \frac{16}
          {15} ) )\,
      \Gamma(-( \frac{11}{15} ) )\,
      {\Gamma(-( \frac{2}{5} ) )}^3}
    ) +\frac{{\Gamma(-( \frac{3}{5} ) )}^2\,
    \Gamma(-( \frac{2}{5} ) )\,
    \Gamma(-( \frac{1}{5} ) )\,
    \Gamma(\frac{1}{5})\,
    \Gamma(\frac{3}{5})}{{( -z ) }^
     {\frac{3}{5}}\,\Gamma(-( \frac{4}{15}
        ) )\,\Gamma(\frac{1}{15})\,
    {\Gamma(\frac{2}{5})}^3}\nonumber\\
& &  -( \frac{{\Gamma(-( \frac{8}{5} )
            )}^2\,\Gamma(-( \frac{7}{5} )
          )\,\Gamma(-( \frac{6}{5} ) )\,
      \Gamma(-( \frac{4}{5} ) )\,
      \Gamma(\frac{8}{5})}{{( -z ) }^
       {\frac{8}{5}}\,\Gamma(-( \frac{19}
          {15} ) )\,
      \Gamma(-( \frac{14}{15} ) )\,
      {\Gamma(-( \frac{3}{5} ) )}^3}
    ) +\frac{{\Gamma(-( \frac{4}{5} ) )}^2\,
    \Gamma(-( \frac{3}{5} ) )\,
    \Gamma(-( \frac{2}{5} ) )\,
    \Gamma(-( \frac{1}{5} ) )\,
    \Gamma(\frac{4}{5})}{{( -z ) }^
     {\frac{4}{5}}\,\Gamma(-( \frac{7}{15}
        ) )\,\Gamma(-( \frac{2}{15}
        ) )\,{\Gamma(\frac{1}{5})}^3}\nonumber\\
& & + -( \frac{{\Gamma(-( \frac{9}{5} )
            )}^2\,\Gamma(-( \frac{8}{5} )
          )\,\Gamma(-( \frac{7}{5} ) )\,
      \Gamma(-( \frac{6}{5} ) )\,
      \Gamma(\frac{9}{5})}{{( -z ) }^
       {\frac{9}{5}}\,\Gamma(-( \frac{22}
          {15} ) )\,
      \Gamma(-( \frac{17}{15} ) )\,
      {\Gamma(-( \frac{4}{5} ) )}^3}
    )+...\biggr]\theta(|z|-1)\nonumber\\
& & +\biggl[\frac{{\cal Y} }
    {2\,\Gamma(\frac{1}{3})\,
    \Gamma(\frac{2}{3})}-( \frac{z\,\Gamma(\frac{6}{5})\,
      \Gamma(\frac{7}{5})\,
      \Gamma(\frac{8}{5})\,
      \Gamma(\frac{9}{5})}{\Gamma(
       \frac{4}{3})\,\Gamma(\frac{5}{3})} )
-  \frac{( z^2\,\Gamma(\frac{11}{5})\,
      \Gamma(\frac{12}{5})\,
      \Gamma(\frac{13}{5})\,
      \Gamma(\frac{14}{5}) ) }{16\,
    \Gamma(\frac{7}{3})\,
    \Gamma(\frac{8}{3})}+...\biggr]\theta(1-|z|),
\end{eqnarray}
where
\begin{eqnarray*}
& & {\cal Y}\equiv -( \Gamma(\frac{1}{5})\,
      \Gamma(\frac{2}{5})\,
      \Gamma(\frac{3}{5})\,
      \Gamma(\frac{4}{5})\,
      ( 4\,{\gamma}^2 + ({\log (-z)})^2 +
        {\psi(\frac{1}{5})}^2 -
        2\,\psi(\frac{1}{5})\,
         \psi(\frac{1}{3}) +
        {\psi(\frac{1}{3})}^2 +
        2\,\psi(\frac{1}{5})\,
         \psi(\frac{2}{5}) -
        2\,\psi(\frac{1}{3})\,
         \psi(\frac{2}{5})\nonumber\\
& &  + {\psi(\frac{2}{5})}^2+
        2\,\psi(\frac{1}{5})\,
         \psi(\frac{3}{5}) -
        2\,\psi(\frac{1}{3})\,
         \psi(\frac{3}{5}) +
        2\,\psi(\frac{2}{5})\,
         \psi(\frac{3}{5}) +
        {\psi(\frac{3}{5})}^2 -
        2\,\psi(\frac{1}{5})\,
         \psi(\frac{2}{3}) +
        2\,\psi(\frac{1}{3})\,
         \psi(\frac{2}{3}) -
        2\,\psi(\frac{2}{5})\,
         \psi(\frac{2}{3})\nonumber\\
& &  -
        2\,\psi(\frac{3}{5})\,
         \psi(\frac{2}{3}) +
        {\psi(\frac{2}{3})}^2+
        2\,\psi(\frac{1}{5})\,
         \psi(\frac{4}{5}) -
        2\,\psi(\frac{1}{3})\,
         \psi(\frac{4}{5}) +
        2\,\psi(\frac{2}{5})\,
         \psi(\frac{4}{5}) +
        2\,\psi(\frac{3}{5})\,
         \psi(\frac{4}{5}) -
        2\,\psi(\frac{2}{3})\,
         \psi(\frac{4}{5}) +
        {\psi(\frac{4}{5})}^2\nonumber\\
& &  +
        4\,\gamma\,
         ( \psi(\frac{1}{5})- \psi(\frac{1}{3}) +
           \psi(\frac{2}{5}) +
           \psi(\frac{3}{5})
- \psi(\frac{2}{3}) +
           \psi(\frac{4}{5}) )  +
        2\,\log (-z)\,( 2\,\gamma+
           \psi(\frac{1}{5}) -
           \psi(\frac{1}{3}) +
           \psi(\frac{2}{5}) +
           \psi(\frac{3}{5}) -
           \psi(\frac{2}{3})\nonumber\\
& &  +
           \psi(\frac{4}{5}) )  +
        \psi^\prime(\frac{1}{5}) -
        \psi^\prime(\frac{1}{3}) +
        \psi^\prime(\frac{2}{5}) +
        \psi^\prime(\frac{3}{5}) -
        \psi^\prime(\frac{2}{3}) +
        \psi^\prime(\frac{4}{5}) )  );
\end{eqnarray*}

\begin{eqnarray}
\label{eq:F2}
& & F_2= \biggl[\frac{{\Gamma(-( \frac{2}{5} ) )}^2\,
    \Gamma(\frac{1}{5})\,
    {\Gamma(\frac{2}{5})}^2}{z^{\frac{2}{5}}\,
    \Gamma(-( \frac{1}{15} ) )\,
    \Gamma(\frac{4}{15})\,
    {\Gamma(\frac{3}{5})}^3\,
    \Gamma(\frac{6}{5})} \frac{{\Gamma(-( \frac{3}{5} ) )}^2\,
    \Gamma(-( \frac{1}{5} ) )\,
    \Gamma(\frac{1}{5})\,
    \Gamma(\frac{3}{5})}{z^{\frac{3}{5}}\,
    \Gamma(-( \frac{4}{15} ) )\,
    \Gamma(\frac{1}{15})\,
    {\Gamma(\frac{2}{5})}^3\,
    \Gamma(\frac{7}{5})}+\frac{{\Gamma(-( \frac{4}{5} ) )}^2\,
    \Gamma(-( \frac{2}{5} ) )\,
    \Gamma(-( \frac{1}{5} ) )\,
    \Gamma(\frac{4}{5})}{z^{\frac{4}{5}}\,
    \Gamma(-( \frac{7}{15} ) )\,
    \Gamma(-( \frac{2}{15} ) )\,
    {\Gamma(\frac{1}{5})}^3\,
    \Gamma(\frac{8}{5})}\nonumber\\
& & -( \frac{{\Gamma(-( \frac{7}{5} )
            )}^2\,\Gamma(-( \frac{4}{5} )
          )\,\Gamma(-( \frac{3}{5} ) )\,
      \Gamma(\frac{7}{5})}{z^{\frac{7}{5}}\,
      \Gamma(-( \frac{16}{15} ) )\,
      \Gamma(-( \frac{11}{15} ) )\,
      {\Gamma(-( \frac{2}{5} ) )}^3\,
      \Gamma(\frac{11}{5})} ) -( \frac{{\Gamma(-( \frac{8}{5} )
            )}^2\,\Gamma(-( \frac{6}{5} )
          )\,\Gamma(-( \frac{4}{5} ) )\,
      \Gamma(\frac{8}{5})}{z^{\frac{8}{5}}\,
      \Gamma(-( \frac{19}{15} ) )\,
      \Gamma(-( \frac{14}{15} ) )\,
      {\Gamma(-( \frac{3}{5} ) )}^3\,
      \Gamma(\frac{12}{5})} )\nonumber\\
& & -( \frac{{\Gamma(-( \frac{9}{5} )
            )}^2\,\Gamma(-( \frac{7}{5} )
          )\,\Gamma(-( \frac{6}{5} ) )\,
      \Gamma(\frac{9}{5})}{z^{\frac{9}{5}}\,
      \Gamma(-( \frac{22}{15} ) )\,
      \Gamma(-( \frac{17}{15} ) )\,
      {\Gamma(-( \frac{4}{5} ) )}^3\,
      \Gamma(\frac{13}{5})} )+...\biggr]\theta(|z|-1)\nonumber\\
& & +\biggl[\frac{ {\cal Z}}
    {2\,\Gamma(\frac{1}{3})\,
    \Gamma(\frac{2}{3})}+\frac{z\,\Gamma(\frac{7}{5})\,
    \Gamma(\frac{8}{5})\,
    \Gamma(\frac{9}{5})}{\Gamma(
     -( \frac{1}{5} ) )\,
    \Gamma(\frac{4}{3})\,
    \Gamma(\frac{5}{3})}+\frac{-( z^2\,\Gamma(\frac{12}{5})\,
      \Gamma(\frac{13}{5})\,
      \Gamma(\frac{14}{5}) ) }{16\,
    \Gamma(-( \frac{6}{5} ) )\,
    \Gamma(\frac{7}{3})\,
    \Gamma(\frac{8}{3})}+...\biggr]\theta(1-|z|),
\end{eqnarray}
where
\begin{eqnarray*}
& & {\cal Z}\equiv -( \Gamma(\frac{2}{5})\,
      \Gamma(\frac{3}{5})\,
      ( 4\,{\gamma}^2 + ({\log (z)})^2 +
        {\psi(\frac{1}{3})}^2 -
        2\,\psi(\frac{1}{3})\,
         \psi(\frac{2}{5}) +
        {\psi(\frac{2}{5})}^2 -
        2\,\psi(\frac{1}{3})\,
         \psi(\frac{3}{5}) +
        2\,\psi(\frac{2}{5})\,
         \psi(\frac{3}{5})\nonumber\\
& &  +
        {\psi(\frac{3}{5})}^2 +
        2\,\psi(\frac{1}{3})\,
         \psi(\frac{2}{3}) -
        2\,\psi(\frac{2}{5})\,
         \psi(\frac{2}{3}) -
        2\,\psi(\frac{3}{5})\,
         \psi(\frac{2}{3}) +
        {\psi(\frac{2}{3})}^2 -
        4\,\gamma\,
         ( \psi(\frac{1}{3}) -
           \psi(\frac{2}{5}) -
           \psi(\frac{3}{5}) +
           \psi(\frac{2}{3}) -
           2\,\psi(\frac{4}{5}) )\nonumber\\
& &   -
        4\,\psi(\frac{1}{3})\,
         \psi(\frac{4}{5}) +
        4\,\psi(\frac{2}{5})\,
         \psi(\frac{4}{5}) +
        4\,\psi(\frac{3}{5})\,
         \psi(\frac{4}{5})-
        4\,\psi(\frac{2}{3})\,
         \psi(\frac{4}{5}) +
        4\,{\psi(\frac{4}{5})}^2 \nonumber\\
& &  + 2\,\log (z)\,( 2\,\gamma -
           \psi(\frac{1}{3}) +
           \psi(\frac{2}{5}) +
           \psi(\frac{3}{5}) -
           \psi(\frac{2}{3}) +
           2\,\psi(\frac{4}{5}) )  -
        \psi^\prime(\frac{1}{3})+
        \psi^\prime(\frac{2}{5}) +
        \psi^\prime(\frac{3}{5}) -
        \psi^\prime(\frac{2}{3}) )  );
\end{eqnarray*}

\begin{eqnarray}
\label{eq:Z_0}
& & Z_0=-( \frac{\Gamma(\frac{1}{5})\,
      \Gamma(\frac{2}{5})\,
      \Gamma(\frac{3}{5})\,
      \Gamma(\frac{4}{5})}{\Gamma(
       \frac{1}{3})\,\Gamma(\frac{2}{3})} )\theta(1-|z|) + \biggl[\frac{{\Gamma(\frac{1}{5})}^2\,
    \Gamma(\frac{2}{5})\,
    \Gamma(\frac{3}{5})}{{( -z ) }^
     {\frac{1}{5}}\,\Gamma(\frac{2}{15})\,
    \Gamma(\frac{7}{15})\,
    {\Gamma(\frac{4}{5})}^3\,
    {\Gamma(\frac{6}{5})}^2}\nonumber\\
& & + \frac{\Gamma(-( \frac{1}{5} ) )\,
    \Gamma(\frac{1}{5})\,
    {\Gamma(\frac{2}{5})}^2}{{( -z ) }^
     {\frac{2}{5}}\,\Gamma(-( \frac{1}{15}
        ) )\,\Gamma(\frac{4}{15})\,
    {\Gamma(\frac{3}{5})}^3\,
    {\Gamma(\frac{7}{5})}^2} + \frac{\Gamma(-( \frac{2}{5} ) )\,
    \Gamma(-( \frac{1}{5} ) )\,
    \Gamma(\frac{1}{5})\,
    \Gamma(\frac{3}{5})}{{( -z ) }^
     {\frac{3}{5}}\,\Gamma(-( \frac{4}{15}
        ) )\,\Gamma(\frac{1}{15})\,
    {\Gamma(\frac{2}{5})}^3\,
    {\Gamma(\frac{8}{5})}^2}\nonumber\\
& & \frac{\Gamma(-( \frac{3}{5} ) )\,
    \Gamma(-( \frac{2}{5} ) )\,
    \Gamma(-( \frac{1}{5} ) )\,
    \Gamma(\frac{4}{5})}{{( -z ) }^
     {\frac{4}{5}}\,\Gamma(-( \frac{7}{15}
        ) )\,\Gamma(-( \frac{2}{15}
        ) )\,{\Gamma(\frac{1}{5})}^3\,
    {\Gamma(\frac{9}{5})}^2} -( \frac{\Gamma(-( \frac{4}{5} ) )\,
      \Gamma(-( \frac{3}{5} ) )\,
      \Gamma(-( \frac{2}{5} ) )\,
      \Gamma(\frac{6}{5})}{{( -z ) }^
       {\frac{6}{5}}\,\Gamma(-( \frac{13}
          {15} ) )\,
      \Gamma(-( \frac{8}{15} ) )\,
      {\Gamma(-( \frac{1}{5} ) )}^3\,
      {\Gamma(\frac{11}{5})}^2} )\nonumber\\
& & + -( \frac{\Gamma(-( \frac{6}{5} ) )\,
      \Gamma(-( \frac{4}{5} ) )\,
      \Gamma(-( \frac{3}{5} ) )\,
      \Gamma(\frac{7}{5})}{{( -z ) }^
       {\frac{7}{5}}\,\Gamma(-( \frac{16}
          {15} ) )\,
      \Gamma(-( \frac{11}{15} ) )\,
      {\Gamma(-( \frac{2}{5} ) )}^3\,
      {\Gamma(\frac{12}{5})}^2} ) + -( \frac{\Gamma(-( \frac{7}{5} ) )\,
      \Gamma(-( \frac{6}{5} ) )\,
      \Gamma(-( \frac{4}{5} ) )\,
      \Gamma(\frac{8}{5})}{{( -z ) }^
       {\frac{8}{5}}\,\Gamma(-( \frac{19}
          {15} ) )\,
      \Gamma(-( \frac{14}{15} ) )\,
      {\Gamma(-( \frac{3}{5} ) )}^3\,
      {\Gamma(\frac{13}{5})}^2} )\nonumber\\
& & -( \frac{\Gamma(-( \frac{8}{5} ) )\,
      \Gamma(-( \frac{7}{5} ) )\,
      \Gamma(-( \frac{6}{5} ) )\,
      \Gamma(\frac{9}{5})}{{( -z ) }^
       {\frac{9}{5}}\,\Gamma(-( \frac{22}
          {15} ) )\,
      \Gamma(-( \frac{17}{15} ) )\,
      {\Gamma(-( \frac{4}{5} ) )}^3\,
      {\Gamma(\frac{14}{5})}^2} )+...\biggr]\theta(|z|-1);
\end{eqnarray}

\begin{eqnarray}
\label{eq:Z_1}
& & \!\!\!\!\!\!\!\!\!\!\!\!\!Z_1=-( \frac{\Gamma(\frac{2}{5})\,
      \Gamma(\frac{3}{5})\,
      ( 2\,\gamma + \log (-z) -
        \psi(\frac{1}{3}) +
        \psi(\frac{2}{5}) +
        \psi(\frac{3}{5}) -
        \psi(\frac{2}{3}) +
        2\,\psi(\frac{4}{5}) ) }{
      \Gamma(\frac{1}{3})\,
      \Gamma(\frac{2}{3})} )\theta(1-|z|)\nonumber\\
      & &  +\biggl[\frac{\Gamma(-( \frac{2}{5} ) )\,
    \Gamma(\frac{1}{5})\,
    {\Gamma(\frac{2}{5})}^2}{{( -z ) }^
     {\frac{2}{5}}\,\Gamma(-( \frac{1}{15}
        ) )\,\Gamma(\frac{4}{15})\,
    {\Gamma(\frac{3}{5})}^3\,
    \Gamma(\frac{6}{5})\,
    \Gamma(\frac{7}{5})}\nonumber\\
& & +\frac{\Gamma(-( \frac{3}{5} ) )\,
    \Gamma(-( \frac{1}{5} ) )\,
    \Gamma(\frac{1}{5})\,
    \Gamma(\frac{3}{5})}{{( -z ) }^
     {\frac{3}{5}}\,\Gamma(-( \frac{4}{15}
        ) )\,\Gamma(\frac{1}{15})\,
    {\Gamma(\frac{2}{5})}^3\,
    \Gamma(\frac{7}{5})\,
    \Gamma(\frac{8}{5})} +\frac{\Gamma(-( \frac{4}{5} ) )\,
    \Gamma(-( \frac{2}{5} ) )\,
    \Gamma(-( \frac{1}{5} ) )\,
    \Gamma(\frac{4}{5})}{{( -z ) }^
     {\frac{4}{5}}\,\Gamma(-( \frac{7}{15}
        ) )\,\Gamma(-( \frac{2}{15}
        ) )\,{\Gamma(\frac{1}{5})}^3\,
    \Gamma(\frac{8}{5})\,
    \Gamma(\frac{9}{5})}\nonumber\\
& & -( \frac{\Gamma(-( \frac{7}{5} ) )\,
      \Gamma(-( \frac{4}{5} ) )\,
      \Gamma(-( \frac{3}{5} ) )\,
      \Gamma(\frac{7}{5})}{{( -z ) }^
       {\frac{7}{5}}\,\Gamma(-( \frac{16}
          {15} ) )\,
      \Gamma(-( \frac{11}{15} ) )\,
      {\Gamma(-( \frac{2}{5} ) )}^3\,
      \Gamma(\frac{11}{5})\,
      \Gamma(\frac{12}{5})} )-( \frac{\Gamma(-( \frac{8}{5} ) )\,
      \Gamma(-( \frac{6}{5} ) )\,
      \Gamma(-( \frac{4}{5} ) )\,
      \Gamma(\frac{8}{5})}{{( -z ) }^
       {\frac{8}{5}}\,\Gamma(-( \frac{19}
          {15} ) )\,
      \Gamma(-( \frac{14}{15} ) )\,
      {\Gamma(-( \frac{3}{5} ) )}^3\,
      \Gamma(\frac{12}{5})\,
      \Gamma(\frac{13}{5})} )\nonumber\\
& & -( \frac{\Gamma(-( \frac{9}{5} ) )\,
      \Gamma(-( \frac{7}{5} ) )\,
      \Gamma(-( \frac{6}{5} ) )\,
      \Gamma(\frac{9}{5})}{{( -z ) }^
       {\frac{9}{5}}\,\Gamma(-( \frac{22}
          {15} ) )\,
      \Gamma(-( \frac{17}{15} ) )\,
      {\Gamma(-( \frac{4}{5} ) )}^3\,
      \Gamma(\frac{13}{5})\,
      \Gamma(\frac{14}{5})})+...\biggr]\theta(|z|-1);\nonumber\\
& &
\end{eqnarray}
\begin{eqnarray}
\label{eq:Z_2} & & \!\!\!\!\!\!Z_2=\frac{-2\,\gamma - \log (-z) +
    \psi(\frac{1}{3}) -
    2\,\psi(\frac{3}{5}) +
    \psi(\frac{2}{3}) -
    2\,\psi(\frac{4}{5})}{\Gamma(
     \frac{1}{3})\,\Gamma(\frac{2}{3})}\theta(1-|z|)
 + \biggl[\frac{\Gamma(-( \frac{3}{5} ) )\,
    \Gamma(\frac{1}{5})\,
    \Gamma(\frac{3}{5})}{{( -z ) }^
     {\frac{3}{5}}\,\Gamma(-( \frac{4}{15}
        ) )\,\Gamma(\frac{1}{15})\,
    {\Gamma(\frac{2}{5})}^3\,
    \Gamma(\frac{6}{5})\,
    \Gamma(\frac{7}{5})\,
    \Gamma(\frac{8}{5})}\nonumber\\
& & + \frac{\Gamma(-( \frac{4}{5} ) )\,
    \Gamma(-( \frac{1}{5} ) )\,
    \Gamma(\frac{4}{5})}{{( -z ) }^
     {\frac{4}{5}}\,\Gamma(-( \frac{7}{15}
        ) )\,\Gamma(-( \frac{2}{15}
        ) )\,{\Gamma(\frac{1}{5})}^3\,
    \Gamma(\frac{7}{5})\,
    \Gamma(\frac{8}{5})\,
    \Gamma(\frac{9}{5})}-( \frac{\Gamma(-( \frac{8}{5} ) )\,
      \Gamma(-( \frac{4}{5} ) )\,
      \Gamma(\frac{8}{5})}{{( -z ) }^
       {\frac{8}{5}}\,\Gamma(-( \frac{19}
          {15} ) )\,
      \Gamma(-( \frac{14}{15} ) )\,
      {\Gamma(-( \frac{3}{5} ) )}^3\,
      \Gamma(\frac{11}{5})\,
      \Gamma(\frac{12}{5})\,
      \Gamma(\frac{13}{5})} )\nonumber\\
& & -( \frac{\Gamma(-( \frac{9}{5} ) )\,
      \Gamma(-( \frac{6}{5} ) )\,
      \Gamma(\frac{9}{5})}{{( -z ) }^
       {\frac{9}{5}}\,\Gamma(-( \frac{22}
          {15} ) )\,
      \Gamma(-( \frac{17}{15} ) )\,
      {\Gamma(-( \frac{4}{5} ) )}^3\,
      \Gamma(\frac{12}{5})\,
      \Gamma(\frac{13}{5})\,
      \Gamma(\frac{14}{5})} )+...\biggr]\theta(|z|-1).
\end{eqnarray}
Analogous to bosonic manifolds, one can make predictions about the
world-sheet instanton contributions to the periods on the super
gauged linear sigma model using the above results for the mirror
Landau-Ginsburg model.

\subsection{Monodromies}

We now discuss monodromy for the mirror super Landau-Ginsburg model
corresponding to the supermanifold considered in this paper. To
discuss the same, the Picard-Fuchs equation can be written in the
form\cite{GL,AM}:
\begin{equation}
\label{eq:PFform}
\biggl(\Delta_z^{6}+\sum_{i=1}^{5}{\bf B}_i(z)\Delta_z^i\biggr){\Pi}(z)
=0.
\end{equation}
The Picard-Fuchs
 equation in the form written in (\ref{eq:PFform}) can alternatively be expressed as
the following system of six linear differential equations:
\begin{eqnarray}
\label{eq:PFdiffeq3} & & \Delta_z\left(\begin{array}{c}
\tilde{\Pi}(z)\\
\Delta_z
\tilde{\Pi}(z)\\
(\Delta_z)^2
\tilde{\Pi}(z)\\
...\\
(\Delta_z)^5
\tilde{\Pi}(z)\\
\end{array}\right)
=\nonumber\\
& & \left(\begin{array}{ccccc}
0 & 1 & 0 & ...  & 0\\
0 & 0 & 1 & ...  & 0\\
. & . & . & ... . & . \\
0 & 0 & 0 & ...  & 1 \\
0 & -{\bf B}_1(z) & -{\bf B}_2(z) & ...  & -{\bf B}_5(z)\\
\end{array}\right)
\left(\begin{array}{c}
\tilde{\Pi}(z)\\
\Delta_z
\tilde{\Pi}(z)\\
(\Delta_z)^2
\tilde{\Pi}(z)\\
...\\
(\Delta_z)^5
\tilde{\Pi}(z)\\
\end{array}\right)
\end{eqnarray}
The matrix on the RHS of (\ref{eq:PFdiffeq3}) is usually denoted by $A(z)$.

 If the
six solutions, ${\Pi}_{I=1,...,6}\}$,
 are collected as a column vector ${\Pi}(z)$, then the {\it constant}
monodromy matrix $T$ for $|z|<<1$ is defined by:
\begin{equation}
{\Pi}(e^{2\pi i}z)=T{\Pi}(z).
\end{equation}
The basis for the space of solutions can be collected as the
columns of the ``fundamental matrix" $\Phi(z)$ given by:
\begin{equation}
\label{eq:PFdiffeqsol}
\Phi(z)=
S_6(z)z^{R_6},
\end{equation}
where $S_6(z)$ and $R_6$ are 6$\times$6 matrices that single and
multiple-valued respectively. Note that ${\bf B}_i(0)\neq0$, which influences
the monodromy properties. Also,
\begin{equation}
\label{eq:Phidieuf2} \Phi(z)=\left(\begin{array}{ccc}
\tilde{\Pi}_1(z) & ... & \tilde{\Pi}_6(z) \\
\Delta_z \tilde{\Pi}_1(z) & ... & \Delta \tilde{\Pi}_6(z) \\
\Delta_z^2 \tilde{\Pi}_2(z) & ... & \Delta^2 \tilde{\Pi}_6(z) \\
... & ... & ... \\
\Delta_z^5 \tilde{\Pi}_1(z) & ... & \Delta_z^5 \tilde{\Pi}_6(z)\\
\end{array}\right),
\end{equation}
implying that
\begin{equation}
\label{eq:Tdieuf}
T=e^{2\pi i R^t}.
\end{equation}
Now, writing $z^R=e^{R ln z}=1 + R ln z + R^2(ln z)^2+...$, and further
noting that there are no terms of order higher than $(ln z)^4$ in
$\tilde{\Pi}(z)$
obtained above, implies that the matrix $R$ must satisfy the property:
$R^{m}=0,\ m=4,...\infty$. Hence, $T=e^{2\pi i R^t}=1 + 2\pi i R^t +
{(2\pi i)^2\over2}(R^t)^2+{(2\pi i)^3\over 6}(R^t)^3.$
Irrespective of whether or not the
distinct eigenvalues of $A(0)$ differ by integers, one has to evaluate
$e^{2\pi i A(0)}$. The eigenvalues of $A(0)$ of (\ref{eq:A(0)}), are $0^4,{1\over3},
{2\over3}$, and hence four of the six eigenvalues differ by an integer (0).

Now, the Picard-Fuchs equation (\ref{eq:SPF3}) can be rewritten in
the form (\ref{eq:PFform}), with the following values of $B_i$'s:
\begin{eqnarray}
\label{eq:Bis}
& & {\bf B}_{1}=0,\ {\bf B}_2=\frac{-24z}{625(1-z)},\
{\bf B}_3={-2z\over5(1-z)},\ {\bf B}_4={(\frac{2}{9}
- \frac{7z}{5})\over(1-z)},\
{\bf B}_5={(-1-2z)\over(1-z)}.\nonumber\\
\end{eqnarray}

Under the change of basis
$\tilde{\Pi}(z)\rightarrow \tilde{\Pi}^\prime(z)=M^{-1}\tilde{\Pi}(z)$,
and writing $\tilde{\Pi}_j(z)=\sum_{i=0}^3(ln z)^iq_{ij}(z)$ (See \cite{GL}), one sees that
\begin{eqnarray}
\label{eq:primed}
& & \tilde{\Pi}^\prime_j(z)=\sum_{i=0}^3(ln z)^iq^\prime_{ij}(z),\nonumber\\
& & q^\prime(z)=q(z)(M^{-1})^t,\nonumber\\
& &  \Phi^\prime(z)=\Phi(z)(M^{-1})^t,\ S^\prime(z)=S(z)(M^{-1})^t,\
R^\prime=M^tR(M^{-1})^t.
\end{eqnarray}
By choosing $M$ such that
$S^\prime(0)={\bf 1}_6$, one gets
\begin{equation}
\label{eq:monodromy1}
T(0)=M(e^{2i\pi A(0)})^tM^{-1}.
\end{equation}
The matrix $A(0)$ is given by:
\begin{equation}
\label{eq:A(0)} A(0)=\left(\begin{array}{cccccccc}
0&1&0&0&0&0\\
0&0&1&0&0&0\\
0&0&0&1&0&0\\
0&0&0&0&1&0\\
0&0&0&0&0&1\\
0&0&0&0&\frac{-2}{9}&1\\
\end{array}\right)
\end{equation}
One can show, using Mathematica, that:
$$e^{2i\pi A(0)}=\hfill$$
$$\!\!\!\!\!\!\!\!\left(\matrix{ 1 & 2\,i\,\pi  & 2\,i^2\,{\pi }^2 & \frac{4\,i^3\,{\pi }^3}{3} & \!\!\!\!\!\frac{-3\,
     \left( 837 - 864\,e^{\frac{2\,i\,\pi }{3}} + 27\,e^{\frac{4\,i\,\pi }{3}} + 540\,i\,\pi  + 168\,i^2\,{\pi }^2 +
       32\,i^3\,{\pi }^3 \right) }{16} &\!\!\!\! \frac{3\,\left( 1215 - 1296\,e^{\frac{2\,i\,\pi }{3}} + 81\,e^{\frac{4\,i\,\pi }{3}} +
       756\,i\,\pi  + 216\,i^2\,{\pi }^2 + 32\,i^3\,{\pi }^3 \right) }{16} \cr 0 & 1 & 2\,i\,\pi  & 2\,i^2\,{\pi }^2 & \frac{-9\,
     \left( 45 - 48\,e^{\frac{2\,i\,\pi }{3}} + 3\,e^{\frac{4\,i\,\pi }{3}} + 28\,i\,\pi  + 8\,i^2\,{\pi }^2 \right) }{8} & \!\!\!\!\!\!
    \frac{9\,\left( 63 - 72\,e^{\frac{2\,i\,\pi }{3}} + 9\,e^{\frac{4\,i\,\pi }{3}} + 36\,i\,\pi  + 8\,i^2\,{\pi }^2 \right) }{8}
   \cr
 0 & 0 & 1 & 2\,i\,\pi  & \frac{-9\,\left( 7 - 8\,e^{\frac{2\,i\,\pi }{3}} + e^{\frac{4\,i\,\pi }{3}} + 4\,i\,\pi  \right) }
   {4} & \frac{9\,\left( 9 - 12\,e^{\frac{2\,i\,\pi }{3}} + 3\,e^{\frac{4\,i\,\pi }{3}} + 4\,i\,\pi  \right) }
   {4} \cr 0 & 0 & 0 & 1 & \frac{-3\,\left( 3 - 4\,e^{\frac{2\,i\,\pi }{3}} + e^{\frac{4\,i\,\pi }{3}} \right) }{2} & \frac{9\,
     {\left( -1 + e^{\frac{2\,i\,\pi }{3}} \right) }^2}{2} \cr 0 & 0 & 0 & 0 & -\left( e^{\frac{2\,i\,\pi }{3}}\,
     \left( -2 + e^{\frac{2\,i\,\pi }{3}} \right)  \right)  & 3\,e^{\frac{2\,i\,\pi }{3}}\,
   \left( -1 + e^{\frac{2\,i\,\pi }{3}} \right)  \cr 0 & 0 & 0 & 0 & \frac{-2\,e^{\frac{2\,i\,\pi }{3}}\,
     \left( -1 + e^{\frac{2\,i\,\pi }{3}} \right) }{3} & e^{\frac{2\,i\,\pi }{3}}\,
   \left( -1 + 2\,e^{\frac{2\,i\,\pi }{3}} \right)  \cr  }\right)$$
Writing the solution vector $\tilde{\Pi}_i$ as $\tilde{\Pi}_i=\sum_{j=0}^4(ln z)^j q_{ji}$
(following the notation of \cite{GL}), one notes:
\begin{equation}
\label{eq:primedsol1}
(\Phi^\prime)_{0i}=(\tilde{\Pi}^\prime)^t_i=\biggl(S^\prime z^{A(0)}\biggr)_{0i}=
(ln z)^j q^\prime_{ji}.
\end{equation}
From (\ref{eq:primedsol1}), one gets the following:
\begin{equation}
\label{eq:primedsol2}
(q^\prime(0))_{ji}={\delta_{ji}\over j!},\ 0\leq (i,j)\leq 3.
\end{equation}
Now, using Mathematica, one gets:
$$z^{A(0)}=\hfill$$
$$\!\!\!\!\!\!\!\!\!
\left(\matrix{ 1 & \log (z) & \!\!\!\frac{{\log (z)}^2}{2} &\!\!\! \frac{{\log (z)}^3}{6} &\!\!\!\!\!\! \frac{-3\,
     \left( 27\,\left( 31 - 32\,z^{\frac{1}{3}} + z^{\frac{2}{3}} \right)  + 270\,\log (z) + 42\,{\log (z)}^2 + 4\,{\log (z)}^3
       \right) }{16} &\!\! \!\!\!\frac{3\,\left( 81\,\left( 15 - 16\,z^{\frac{1}{3}} + z^{\frac{2}{3}} \right)  + 378\,\log (z) +
       54\,{\log (z)}^2 + 4\,{\log (z)}^3 \right) }{16} \cr 0 & 1 & \log (z) & \frac{{\log (z)}^2}{2} & \frac{-9\,
     \left( 45 - 48\,z^{\frac{1}{3}} + 3\,z^{\frac{2}{3}} + 14\,\log (z) + 2\,{\log (z)}^2 \right) }{8} & \frac{9\,
     \left( 9\,\left( 7 - 8\,z^{\frac{1}{3}} + z^{\frac{2}{3}} \right)  + 18\,\log (z) + 2\,{\log (z)}^2 \right) }{8} \cr 0 & 0 &
   1 & \log (z) & \frac{-9\,\left( 7 - 8\,z^{\frac{1}{3}} + z^{\frac{2}{3}} + 2\,\log (z) \right) }{4} & \frac{9\,
     \left( 3\,\left( 3 - 4\,z^{\frac{1}{3}} + z^{\frac{2}{3}} \right)  + 2\,\log (z) \right) }{4} \cr 0 & 0 & 0 & 1 & \frac{-3\,
     \left( 3 - 4\,z^{\frac{1}{3}} + z^{\frac{2}{3}} \right) }{2} & \frac{9\,{\left( -1 + z^{\frac{1}{3}} \right) }^2}
   {2} \cr 0 & 0 & 0 & 0 & \left( 2 - z^{\frac{1}{3}} \right) \,z^{\frac{1}{3}} & 3\,\left( -1 + z^{\frac{1}{3}} \right) \,
   z^{\frac{1}{3}} \cr 0 & 0 & 0 & 0 & \frac{-2\,\left( -1 + z^{\frac{1}{3}} \right) \,z^{\frac{1}{3}}}{3} &
   \left( -1 + 2\,z^{\frac{1}{3}} \right) \,z^{\frac{1}{3}} \cr  }\!\!\!\right)$$
Then from the expression for $z^{A(0)}$ above, writing
\begin{equation}
\label{eq:zA0} (e^{A(0)})_{0i}=f_{oi}(z^{\frac{1}{3}}) +
\sum_{n=1}^3c^{(0i)}_n(ln z)^n,
\end{equation}
where $0\leq i\leq 5$, one gets:
\begin{eqnarray}
\label{eq:primedsol6}
& & (q^\prime(0))_{0i}=f_{0i}(0),\nonumber\\
& & (q^\prime(0))_{ij}=c_i^{0j},\ 1\leq i\leq 3,\ 4\leq j\leq 5.
\end{eqnarray}
One can show that the matrices $q$ and $q^\prime$ are given as:
$$q=\left(\matrix{q_{00} & q_{01} & q_{02} & q_{03} & q_{04} & q_{05} \cr
q_{10} & q_{11} & q_{12} & 0 &q_{14} & q_{15} \cr
q_{20} & q_{21} & q_{22} & 0 & 0 & 0 \cr
q_{30} & 0 & 0 & 0 & 0 & 0 }\right)\hfill,$$
and
$$q^\prime=\left(\matrix{ 1 & 0 & 0 & 0 & -\left( \frac{81}{16} \right)  & \frac{243}{16} \cr 0 & 1 & 0 & 0 & -\left( \frac{405}{8} \right)  &
    \frac{567}{8} \cr 0 & 0 & \frac{1}{2} & 0 & -\left( \frac{63}{8} \right)  & \frac{81}{8} \cr 0 & 0 & 0 & \frac{1}{6} & -\left(
     \frac{3}{4} \right)  & \frac{3}{4} \cr  }\right).$$
Finally, using $q^\prime=q(M^{-1}\ ^t)$, one gets 24 equations in 36
elements of $M$. Further constraints on the 36-24=12 elements are
$M$ are expected to come by imposing the requirement $(T^{n}-{\bf
1})^m=0$ for some $n,m\in{\bf Z}^+$ (See \cite{GL} and references
therein).

For $|z|>>1$, the period vector can be written as $\Pi_i=(A_{ij}(\infty)\pi_j$, where $\pi_j\equiv z^{-\frac{j}{5}}$. One thus sees that the monodromy for $u_j$ is given by the matrix
$$T_\pi(\infty)=\left(\matrix{e^{\frac{-2i\pi}{5}} & 0 & 0 & 0 \cr 0 & e^{{-4i\pi\over5}} & 0 & 0 \cr
0 & 0 & e^{\frac{-6i\pi}{5}} & 0\cr 0 & 0 & 0 & e^{\frac{-8 i\pi}{5}}}\right),\hfill$$ using which
the monodromy at $\infty$, $T(\infty)$, of the period vector can be determined from the equation:
\begin{equation}
\label{eq:Monodromyinfty}
A(\infty)T_\pi(\infty)=T(\infty)A(\infty).
\end{equation}
The matrix $A(\infty)$ is given by:
$$A(\infty)=\left(\matrix{A_{11} & A_{12} & A_{13} & A_{14}\cr A_{21} & A_{22} & A_{23} & A_{24} \cr 0 & A_{32} & A_{33} & A_{34} \cr A_{41} & A_{42} & A_{43} & A_{44}\cr 0 & A_{52} & A_{53} & A_{54} \cr 0 & 0 & A_{63} & A_{64}
} \right),$$
where
$$A_{11}=\frac{{\Gamma(-( \frac{1}{5} ) )}^2\,
    {\Gamma(\frac{1}{5})}^3\,
    \Gamma(\frac{2}{5})\,
    \Gamma(\frac{3}{5})}{\,
    \Gamma(\frac{2}{15})\,
    \Gamma(\frac{7}{15})\,
    {\Gamma(\frac{4}{5})}^2},\
A_{12}=\frac{{\Gamma(-( \frac{2}{5} ) )}^2\,
    \Gamma(-( \frac{1}{5} ) )\,
    \Gamma(\frac{1}{5})\,
    {\Gamma(\frac{2}{5})}^3}{\,
    \Gamma(-( \frac{1}{15} ) )\,
    \Gamma(\frac{4}{15})\,
    {\Gamma(\frac{3}{5})}^2},\hfill$$
$$A_{13}= \frac{{\Gamma(-( \frac{3}{5} ) )}^2\,
    \Gamma(-( \frac{2}{5} ) )\,
    \Gamma(-( \frac{1}{5} ) )\,
    \Gamma(\frac{1}{5})\,
    {\Gamma(\frac{3}{5})}^2}{\,
    \Gamma(-( \frac{4}{15} ) )\,
    \Gamma(\frac{1}{15})\,
    {\Gamma(\frac{2}{5})}^2},\
A_{14}= \frac{{\Gamma(-( \frac{4}{5} ) )}^2\,
    \Gamma(-( \frac{3}{5} ) )\,
    \Gamma(-( \frac{2}{5} ) )\,
    \Gamma(-( \frac{1}{5} ) )\,
    {\Gamma(\frac{4}{5})}^2}{\,
    \Gamma(-( \frac{7}{15} ) )\,
    \Gamma(-( \frac{2}{15} ) )\,
    {\Gamma(\frac{1}{5})}^2},\hfill$$
$$A_{21}=\frac{{\Gamma(-( \frac{1}{5} ) )}^2\,
    {\Gamma(\frac{1}{5})}^2\,
    \Gamma(\frac{2}{5})\,
    \Gamma(\frac{3}{5})}{\,\Gamma(\frac{2}{15})\,
    \Gamma(\frac{7}{15})\,
    {\Gamma(\frac{4}{5})}^3},\
A_{22}= -\frac{{\Gamma(-( \frac{2}{5} ) )}^2\,
    \Gamma(-( \frac{1}{5} ) )\,
    \Gamma(\frac{1}{5})\,
    {\Gamma(\frac{2}{5})}^2}{\,\Gamma(-( \frac{1}{15}
        ) )\,\Gamma(\frac{4}{15})\,
    {\Gamma(\frac{3}{5})}^3},\hfill$$
$$A_{23}=\frac{{\Gamma(-( \frac{3}{5} ) )}^2\,
    \Gamma(-( \frac{2}{5} ) )\,
    \Gamma(-( \frac{1}{5} ) )\,
    \Gamma(\frac{1}{5})\,
    \Gamma(\frac{3}{5})}{\,\Gamma(-( \frac{4}{15}
        ) )\,\Gamma(\frac{1}{15})\,
    {\Gamma(\frac{2}{5})}^3},\
A_{24}=\frac{{\Gamma(-( \frac{4}{5} ) )}^2\,
    \Gamma(-( \frac{3}{5} ) )\,
    \Gamma(-( \frac{2}{5} ) )\,
    \Gamma(-( \frac{1}{5} ) )\,
    \Gamma(\frac{4}{5})}{\,\Gamma(-( \frac{7}{15}
        ) )\,\Gamma(-( \frac{2}{15}
        ) )\,{\Gamma(\frac{1}{5})}^3},\hfill$$
$$A_{32}=\frac{{\Gamma(-( \frac{2}{5} ) )}^2\,
    \Gamma(\frac{1}{5})\,
    {\Gamma(\frac{2}{5})}^2}{\,
    \Gamma(-( \frac{1}{15} ) )\,
    \Gamma(\frac{4}{15})\,
    {\Gamma(\frac{3}{5})}^3\,
    \Gamma(\frac{6}{5})},\
A_{33}=\frac{{\Gamma(-( \frac{3}{5} ) )}^2\,
    \Gamma(-( \frac{1}{5} ) )\,
    \Gamma(\frac{1}{5})\,
    \Gamma(\frac{3}{5})}{\,
    \Gamma(-( \frac{4}{15} ) )\,
    \Gamma(\frac{1}{15})\,
    {\Gamma(\frac{2}{5})}},\hfill$$
$$A_{34}=\frac{{\Gamma(-( \frac{4}{5} ) )}^2\,
    \Gamma(-( \frac{2}{5} ) )\,
    \Gamma(-( \frac{1}{5} ) )\,
    \Gamma(\frac{4}{5})}{\,
    \Gamma(-( \frac{7}{15} ) )\,
    \Gamma(-( \frac{2}{15} ) )\,
    {\Gamma(\frac{1}{5})}^3\,
    \Gamma(\frac{8}{5})},\
A_{41}=\frac{{\Gamma(\frac{1}{5})}^2\,
    \Gamma(\frac{2}{5})\,
    \Gamma(\frac{3}{5})}{\,\Gamma(\frac{2}{15})\,
    \Gamma(\frac{7}{15})\,
    {\Gamma(\frac{4}{5})}^3\,
    {\Gamma(\frac{6}{5})}^2},\hfill$$
$$A_{42}=\frac{\Gamma(-( \frac{1}{5} ) )\,
    \Gamma(\frac{1}{5})\,
    {\Gamma(\frac{2}{5})}^2}{\,\Gamma(-( \frac{1}{15}
        ) )\,\Gamma(\frac{4}{15})\,
    {\Gamma(\frac{3}{5})}^3\,
    {\Gamma(\frac{7}{5})}^2},\
A_{43}=\frac{\Gamma(-( \frac{2}{5} ) )\,
    \Gamma(-( \frac{1}{5} ) )\,
    \Gamma(\frac{1}{5})\,
    \Gamma(\frac{3}{5})}{\,\Gamma(-( \frac{4}{15}
        ) )\,\Gamma(\frac{1}{15})\,
    {\Gamma(\frac{2}{5})}^3}\,\hfill$$
$$A_{44}=\frac{\Gamma(-( \frac{3}{5} ) )\,
    \Gamma(-( \frac{2}{5} ) )\,
    \Gamma(-( \frac{1}{5} ) )\,
    \Gamma(\frac{4}{5})}{\,\Gamma(-( \frac{7}{15}
        ) )\,\Gamma(-( \frac{2}{15}
        ) )\,{\Gamma(\frac{1}{5})}^3\,
    {\Gamma(\frac{9}{5})}^2},\
A_{52}=\frac{\Gamma(-( \frac{2}{5} ) )\,
    \Gamma(\frac{1}{5})\,
    {\Gamma(\frac{2}{5})}^2}{\,\Gamma(-( \frac{1}{15}
        ) )\,\Gamma(\frac{4}{15})\,
    {\Gamma(\frac{3}{5})}^3\,
    \Gamma(\frac{6}{5})\,
    \Gamma(\frac{7}{5})},\hfill$$
$$A_{53}=\frac{\Gamma(-( \frac{3}{5} ) )\,
    \Gamma(-( \frac{1}{5} ) )\,
    \Gamma(\frac{1}{5})\,
    \Gamma(\frac{3}{5})}{\,\Gamma(-( \frac{4}{15}
        ) )\,\Gamma(\frac{1}{15})\,
    {\Gamma(\frac{2}{5})}^3\,
    \Gamma(\frac{7}{5})\,
    \Gamma(\frac{8}{5})},\
A_{54}=\frac{\Gamma(-( \frac{4}{5} ) )\,
    \Gamma(-( \frac{2}{5} ) )\,
    \Gamma(-( \frac{1}{5} ) )\,
    \Gamma(\frac{4}{5})}{\,\Gamma(-( \frac{7}{15}
        ) )\,\Gamma(-( \frac{2}{15}
        ) )\,{\Gamma(\frac{1}{5})}^3\,
    \Gamma(\frac{8}{5})\,
    \Gamma(\frac{9}{5})},\hfill$$
$$A_{63}=\frac{\Gamma(-( \frac{3}{5} ) )\,
    \Gamma(\frac{1}{5})\,
    \Gamma(\frac{3}{5})}{\,\Gamma(-( \frac{4}{15}
        ) )\,\Gamma(\frac{1}{15})\,
    {\Gamma(\frac{2}{5})}^3\,
    \Gamma(\frac{6}{5})\,
    \Gamma(\frac{7}{5})\,
    \Gamma(\frac{8}{5})},\
A_{64}=\frac{\Gamma(-( \frac{4}{5} ) )\,
    \Gamma(-( \frac{1}{5} ) )\,
    \Gamma(\frac{4}{5})}{\,\Gamma(-( \frac{7}{15}
        ) )\,\Gamma(-( \frac{2}{15}
        ) )\,{\Gamma(\frac{1}{5})}^3\,
    \Gamma(\frac{7}{5})\,
    \Gamma(\frac{8}{5})\,
    \Gamma(\frac{9}{5})}.$$

Using the argument of \cite{GL}, one sees that the monodromy at 1 is related to the same at 0 and $\infty$
by the relation:
\begin{equation}
\label{eq:monodromy2}
T(1)=\biggl(T(0)\biggr)^{-1}T(\infty).
\end{equation}

\section{The Mirror Hypersurface}

In this section, following \cite{AM}, we obtain the mirror hypersurface to
the super Calabi-Yau in ${\bf WCP}^{(3|2)}[1,1,1,3|1,5]$. To do so, we
first integrate out $X^0$ to yield:
\begin{equation}
\label{eq:hypermirror1}
\int\prod_{i=1}^3 dY^I dX^1 d\eta^1 d\chi^1 dY^0 d\eta^0 d\chi^0 e^{\sum_{J=0}^3e^{-Y^J} + e^{-\sum_{I=1}^3 Q^IY^I + 5X^1 + t} + e^{-\sum_{I=1}^3 Q^IY^I + 5X^1 + t}\eta^0\chi^0 + e^{-X^1}(1+\eta^1\chi^1)},
\end{equation}
which after redefining $e^t\chi^0$ as $\chi^0$ and introducing $\hat{Y}^I$ and $\hat{X}^1$ via the definitions:
$Y^I=\hat{Y}^I+Y^0$ and $X^1=\hat{X}^1+Y^0$, and integrating out $\eta^0$ and $\chi^0$:
\begin{equation}
\label{eq:hypermirror2}
\int\prod_{I=1}^3 dY^IdX^1 d\eta^1 d\chi^1 dY^0 e^{-Y^0-\sum_{i=1}^3Q^i(\hat{Y}^i+Y^0) + 5X^1}
Exp[e^{-Y^0}(1+e^{-\sum_{I=1}^3Q^I\hat{Y}^I+5X^1+t} + \sum_{I=1}^3e^{-Y^I}+e^{-X^1}+e^{-X^1}\eta^1\chi^1)].
\end{equation}
Defining $\Lambda\equiv e^{-Y^0}$, one gets:
\begin{equation}
\int\prod_{I=1}^3 dY^I dX^1 d\eta^1 d\chi^1 e^{\Lambda(1+e^{-\sum_{I=1}^3Q^I\hat{Y}^I+5X^1+t}+\sum_{I=1}^3e^{-Y^I}+e^{-X^1}+e^{-X^1}\eta^1\chi^1)},
\end{equation}
which after the redefinitions:
\begin{equation}
\label{eq:redefs}
e^{-\hat{Y}_1}\equiv (x_1y_1)^6,\ e^{-\hat{Y}_2}\equiv y_2^6,\ e^{-\hat{Y}_3}\equiv y_3^2,\ \ e^{-\hat{X}_1}\equiv x_1^{\frac{6}{5}},
\end{equation}
gives the following mirror hypersurface after performing the $\Lambda$ integral:
\begin{equation}
\label{eq:hypersurface1}
1+e^ty_1^6y_2^6y_3^6+(x_1y_1)^6+y_2^6+y_3^2+x_1^{\frac{6}{5}}+x_1^{\frac{6}{5}}\eta^1\chi^1=0.
\end{equation}
One can rewrite (\ref{eq:hypersurface1}) by defining $x_1\equiv u^5$ and $x_1^{\frac{6}{5}}\eta^1$ as $\eta^1$
as:
\begin{equation}
\label{eq:hypersurface2}
1+e^ty_1^6y_2^6y_3^6+y_1^6u^{30}+y_2^6+y_3^2+u^5+\eta^1\chi^1=0,
\end{equation}
which in the limit $t\rightarrow-\infty$, and appropriately shifting $y_1$,
gives:
\begin{equation}
\label{eq:hypersurface3}
1+y_1^6u^{30}+y_2^6+y_3^2+\eta^1\chi^1=0,
\end{equation}
or
\begin{equation}
\label{eq:hypersurface4}
1+y_1^6x_1^6+y_2^6+y_3^2+\eta^1\chi^1=0.
\end{equation}
Interestingly,
the mirror hypersurface (\ref{eq:hypersurface4}) can be viewed either as

(a) (assuming that the inhomogeneous coordinates $x_1,\chi_1$ and
$y_1,y_2,y_3,\eta_1$ are to be thought of as the following ratio
$\frac{x_1}{x_0},\frac{\chi_1}{x_0^6}; \frac{y_1}{y_0},\frac{y_2}{y_0},
\frac{y_3}{y_0^2},\frac{\eta_1}{y_0^6}$
of homogeneous coordinates $x_0,x_1,\chi_1;y_0,y_1,y_2,y_3$)
as a bidegree-(6,6) hypersurface in
the $x^0=y^0=1$ coordinate patch of the non-singular supermanifold
 ${\bf WCP}^{(3|1)}[1,1,1,2|6](y^{I=0,1,2,3},\eta^1)\times{\bf WCP}^{(1|1)}
[1,1|6](x^{J=0,1},\chi^1)$:
\begin{equation}
\label{hypersurface4}
x_0^6y_0^6 + x_1^6y_1^6 + x_0^6y_2^6 + x_0^6y_0^2y_3^2+\eta_1\chi_1=0;
\end{equation}
{\bf as the sum of the weights for the
bosonic and Grassmanian coordinates do not match, neither
 of the ${\bf WCP}$'s corresponds to a super Calabi-Yau} (See \cite{Sethi}),

or

(b) (assuming that the inhomogeneous coordinates $x_1,\chi_1$ and
$y_1,y_2,y_3,\eta_1$ are to be thought of as the following ratio
$\frac{x_1}{x_0},\frac{\chi_1}{x_0^6}; \frac{y_1}{y_0},\frac{y_2}{y_0^2},
\frac{y_3}{y_0^6},\frac{\eta_1}{y_0^6}$
of homogeneous coordinates $x_0,x_1,\chi_1;y_0,y_1,y_2,y_3$)
as a bidegree-(6,12) hypersurface in the $x^0=y^0=1$ coordinate patch of
the (${\bf Z}_2$) singular supermanifold ${\bf WCP}^{(3|1)}[1,1,2,6|6]\times
{\bf WCP}^{(1|1)}[1,1|6]$:
\begin{equation}
\label{eq:hypersurface5}
y_0^{12}x_0^6 + y_0^6y_1^6x_1^6 + x_0^6y_2^6 + x_0^6y_3^2 + y_0^6\eta_1\chi_1 = 0.
\end{equation}
Once again neither of the supermanifolds
that are multiplied, are super Calabi-Yau's for the same reason as given above.

As part of future work, it will be interesting to verify by calculations of
correlation functions to see either which of the two LG models actually
correspond to the mirror GLSM or whether both are the LG duals but in different
corners of the moduli space - given that the hypersurface in the GLSM side was
non-singular, perhaps it is the non-singular GL dual that will be chosen.
The results of this paper can be readily generalized to other super weighted
 complex projective spaces.


\begin{thebibliography}{99}
\bibitem{AV}M.~Aganagic and C.~Vafa,
{\it Mirror symmetry and supermanifolds},
  arXiv:hep-th/0403192.
\bibitem{HV}K.~Hori and C.~Vafa, {\it Mirror symmetry}, arXiv:hep-th/0002222.
\bibitem{GL}B.~R.~Greene and C.~I.~Lazaroiu,
{\it Collapsing D-branes in Calabi-Yau moduli space. I},
  Nucl.\ Phys.\ B {\bf 604}, 181 (2001)
  [arXiv:hep-th/0001025].
\bibitem{AM}A.~Misra,
{\it Type IIA on a compact Calabi-Yau and D = 11 supergravity uplift of its
orientifold},
  Fortsch.\ Phys.\  {\bf 52}, 831 (2004)
  [arXiv:hep-th/0311186].
\bibitem{Witten}E.~Witten,
{\it Perturbative gauge theory as a string theory in twistor space},
  Commun.\ Math.\ Phys.\  {\bf 252}, 189 (2004)
  [arXiv:hep-th/0312171].
\bibitem{Sethi}S.~Sethi,
{\it Supermanifolds, rigid manifolds and mirror symmetry},
  Nucl.\ Phys.\ B {\bf 430}, 31 (1994)
  [arXiv:hep-th/9404186].
\end{thebibliography}
\end{document}